\documentclass[%
 aps,
prb,%
 amsmath,amssymb,
reprint,%
showpacs,
]{revtex4-1}

\usepackage{graphicx}
\usepackage{dcolumn}
\usepackage{bm}
\usepackage{blkarray}
\usepackage[usenames,dvipsnames]{color}
\usepackage[hyperfootnotes=false]{hyperref}
\usepackage[hyperfootnotes=false]{hyperref}
\usepackage{threeparttable}
\hypersetup{
 colorlinks,
 citecolor=Violet,
 linkcolor=Red,
 filecolor= green,
 urlcolor=Blue}
\begin{document}

\title{Length dependent lattice thermal conductivity of single \& multi layered hexagonal boron nitride: A first-principles study using the Callaway-Klemens \& real space super cell methods}

\author{Ransell D'Souza}%
\email{ransell.d@gmail.com; ransell.dsouza@bose.res.in}

\author{Sugata Mukherjee}%
\email{sugata@bose.res.in; sugatamukh@gmail.com}

\affiliation{Department of Condensed Matter Physics and Materials Science, S.N. Bose National Centre for Basic Sciences, Block JD, Sector III, Salt Lake, Kolkata 700106, India}

\begin{abstract}
The phonon dispersion, density of states, Gr\"{u}neisen parameters, and the lattice thermal conductivity of single- and multi-layered boron nitride were calculated using first-principles methods. For the bulk {\it h}-BN we also report the two-phonon density of states.
We also present simple analytical solutions to the acoustic vibrational mode-dependent lattice thermal conductivity.
Moreover, computations based on the elaborate Callaway-Klemens and the real space super cell methods are presented to calculate the sample length and temperature dependent lattice thermal conductivity of single- and multi-layered hexagonal boron nitride which shows good agreement with experimental data.
\end{abstract}



\maketitle

\section{Introduction}
Single and multilayered boron nitride are $sp^2$ bonded boron and nitrogen atoms arranged in a hexagonal
honeycomb lattice arranged in ABAB stacking in multilayered and bulk materials. In spite of the fact that they are isomorphic to the multilayered graphene and graphite with similar lattice constants, unit cell masses and Van der Waals type bonding between the layers, their phonon properties are quite different. Consequently their physical properties such as the lattice thermal conductivity
 derived from the phonon dispersion should shed light on the fundamental physics of phonon transport of such two-dimensional (2D) nanomaterials. 
 These nanomaterials in the form of semiconductor multilayers and other superstructures are promising candidates of materials with enhanced thermoelectrical properties and have been a topic of intensive research in recent years\cite{duana16}. 
In contrast to a large amount of theoretical and experimental work carried out on electron transport, only few
studies on phonon transport have been reported.
  For example, using density functional theory (DFT) with quantum transport device simulation based on non-equilibrium Green's function (NEGF), Fiori $et\ al$ \cite{fiori11} have proposed and investigated 2-D graphene transistors based on lateral heterobarriers. $Ab\ initio$ atomistic simulations on vertical heterobarrier graphene transistors have been analysed \cite{sciambi11,mehr12}. Britnell $et\ al$ \cite{britnell12} have modelled graphene heterostructures devices  with atomically thin boron nitride as a vertical transport barrier.

Performance of any thermoelectric material is characterised by a dimensionless parameter termed as figure of merit, denoted by $ZT$,
which is inversely proportional to total lattice thermal conductivity, including contributions due to electrons and phonons. However, for the materials investigated here, electron contribution is negligible compared to that of phonons owing to a considerable electronic band gap between their conduction and valence bands.
 Experiments to study the effects of grain-boundaries on the thermal transport properties of graphene have been carried out by a few groups \cite{xu14, ma17}.
These experiments show that a smaller sample length decreases the thermal conductivity, a necessity for a good thermoelectric material.
Graphene has a higher thermal conductivity compared to graphite due to the long mean free path (MFP)
 of the phonons in the 2D lattices. The MFP can thus be reduced by creating defects in the sample. 
 Recent studies by Malekpour {\it el al.} \cite{malekpour16} has shown that vacancies reduces the lattice thermal conductivity in graphene. 
Lattice thermal conductivity of a material is highly correlated to the thickness of the sample (or number of layers). For example, graphene has a much larger thermal conductivity than bilayer graphene and graphite. \cite{RDSM17,hongyang2014}. Recently reported lattice thermal conductivity for In$_2$Se$_3$ exhibits \cite{zhou16} a strong dependence on the thickness or number of layers, with a value of 4 W/mK for a thickness of 5nm which increases to 60 W/mK for the sample with thickness 35nm.
These results suggest that in order to manipulate the lattice thermal conductivity $\kappa_L$, a proper understanding of its dependence on the grain size, temperature and thickness dependence is essential.
However, not many experiments on grain size, temperature and thickness have been carried out so far in single and multilayered boron nitride. We believe our present work will motivate experiments in the direction of tuning $\kappa_L$ in such 2D materials.

Heat flow in single and multilayered boron nitride (SLBN and MLBN) is of great significance not only for fundamental understanding of such materials in terms of lattice thermal conductivity or thermoelectrics but also for technological applications. Single and MLBN are extremely atomically stable materials and can be easily supported between two leads. Besides, these materials exhibit a comparatively lower $\kappa_L$ in bulk than in single and multi-layered graphene. This makes SLBN and MLBN a good testing ground to study the length and temperature dependence of thermal conductivity.
 Manipulating the lattice thermal conductivity by varying its temperature and dimensions (through grain size engineering) will shed light on the fundamental understanding of thermoelectricity in such 2D materials and help in designing new novel materials for technological applications.
 
Hexagonal boron nitride (h-BN) is relatively inert as compared to graphene due to its strong, in-plane, ionic bonding of its planar lattice structure and hence is a favourable substrate dielectric to improve graphene based devices \cite{dean10}. Although h-BN has appealing thermal properties,
  most studies, both experimentally and theoretically,  are confined to single and multi-layered graphene \cite{chen2011,chen2012,balandin08,ghosh08,cai10,jauregui10,hongyang2014,nika2009,RDSM17}.
  Some experiments on lattice thermal conductivity ($\kappa_L$) have been reported by Jo $et\ al$ \cite{jo13} for multi-layered boron nitride
   (MLBN). Also, theoretical studies on thermal conductivity ($\kappa_L$) \cite{lindsay11} and conductance \cite{RDSM16} on such materials have been carried out using Tersoff empirical interatomic potential.
However, first principle theoretical studies of $\kappa_L$ such as using the Boltzmann transport equations (BTE) for phonons from density functional perturbation theory (DFPT) on SLBN and MLBN are apparently not available.

In this paper, we investigate numerically the sample length and temperature dependence of the thermal conductivity ($\kappa_L$) of single and multilayer h-BN 
by solving the phonon BTE beyond the relaxation time approximation (RTA) using the force constant derived from a real space super cell method, and also by solving the phonon BTE in the RTA using the Callaway-Klemens approach.
A long standing puzzle has been to answer which acoustic phonon mode dominates the total lattice thermal conductivity for such 2D materials \cite{nika11}. 
There have been arguments on whether the out-of-plane ZA vibrational mode contributions to $\kappa_L$ are the most dominant or the least in comparison to the other 
acoustic modes. Owing to the selection rules restricting the phase space
for phonon-phonon scattering in ideal graphene \cite{lindsay2010,lindsay10,seol10} and boron nitride \cite{lindsay11}, the ZA mode seem to be the most dominant.
In a rather sharp contrast, references \cite{nika2009,nika09} suggest that since in the long wavelength limit ($q \rightarrow 0$), the phonon dispersion of the ZA modes
seem to be flat thus making the phonon velocities small, and also the fact that the Gr\"{u}neisen parameters are large, would make the ZA contributions to $\kappa_L$ the 
least in comparison to other acoustic modes. 
 Here, using the Callaway-Klemens approach, we examine this discrepancy from analytical solutions to the phonon BTE for each of the acoustic modes using a closed form for the scattering rate for the three-phonon processes derived by Roufosse {\it et. al.} \cite{roufosse73} and an exact numerical solution for the phonon BTE beyond the relaxation time approximation (RTA) in which the phonon lifetimes are formed in terms of a set of coupled equations and solved iteratively.
We also examined the sample length ($L$) dependence of $\kappa_L$ and found this to be very sensitive to $L$, which may justify the application of multilayered h-BN in thermoelectric devices by manipulating $\kappa_L$. 

In the next section we describe the theoretical framework and the first-principles DFT based calculational methods of $\kappa_L$. This is followed by
the results obtained using the real space supercell method and the Callaway-Klemens method and a summary in the subsequent sections.

\section{Theoretical framework and Method of calculation}
\subsection{Electronic and phonon bandstructure calculations} 
First-principles DFT and DFPT calculations were carried out on a hexagonal supercell for the monolayer, bilayer and bulk boron nitride, whereas an orthorhombic supercell was used for five layers h-BN sample, using the plane
wave pseudopotential method as implemented in the QUANTUM ESPRESSO code \cite{giannozzi09}. 
We have used 2 atoms in the unit cell for SLBN, 20 atoms for five layered BN and 4 atoms in both bilayer and bulk boron nitride. 
To prevent interactions between the layers, a vacuum spacing of 20 \AA was introduced along the perpendicular direction to the layers ($z$-axis) mimicking an infinite BN sheet in the $xy$ plane. 
For MLBN and bulk-{\it h}BN, the Van der Waals interaction as prescribed by Grimme \cite{grimme1}, was used between the layers.

For the electronic structure calculations, Monkhorst-Pack grids of $16 \times 16 \times 1$ and $16 \times 16 \times 4$ were chosen for SLBN and MLBN,  resepectively, for the $k$-point sampling. Self-consistent calculations with a 40 Ry kinetic energy cut-off and a 160 Ry
charge density energy cutoff were used to solve the Kohn-Sham equations with an accuracy of 10$^{-9}$ Ry for the total energy. We used ultrasoft pseudopotential to describe the atomic cores with exchange-correlation potential kernel in the local density approximation \cite{rrkj90}. The electronic structure and total enrgy calculations were used to obtain the groundstate geometry before persuing the phonon calculations.

For the phonon bandstructure calculations, the $q$-grid used in the calculations were $6 \times 6 \times 1$ for SLBN,  $6 \times 6 \times 2$ for BLBN and bulk h-BN and 
$4 \times 4 \times 2$ for 5-layer BN, respectively. The density functional perturbation theory (DFPT) \cite{dfpt87}, as implemented in the plane wave method \cite{giannozzi09}, was used to calculate the phonon dispersion and phonon density of states (DOS) along the high-symmetric $q$-points.

\subsection{Calculation of the lattice thermal conductivity}
The calculation of lattice thermal conductivity $\kappa_L$ involves evaluation of second-order harmonic interatomic force constants (IFCs) as well as the third-order anharmonic IFCs. We have used first a real space supercell method which evaluates the third-order IFCs in a real space grid using DFT \cite{ShengBTE}, whereas the second-order IFCs are obtained from the DFPT method \cite{giannozzi09,dfpt87}. Secondly, using the Callaway-Klemens method \cite{callaway59,klemens58}, the relaxation times were obtained from the Gr\"ueisen parameters. Finally, the length, thickness and temperature dependence of $\kappa_L$ were studied.

\subsubsection{Real space super cell approach}
In this method the third order anharmonic IFCs are calculated from a set of displaced supercell configurations depending on the size of the system, their symmetry group and the number of nearest neighbour interactions. 
A $4 \times 4 \times 2$ supercell including upto third nearest neighbour interactions were used to calculate the anharmonic IFCs for all the structures, generating 128 configurations for single and bulk BN, 156 for bilayer BN (BLBN) and 828 for five-layered BN (5LBN). 
The third order anharmonic IFCs are constructed from a set of third-order derivatives of energy, calculated from these configurations using the plane wave method 
\cite{giannozzi09}. The phonon lifetimes are calculated from the phonon BTE which are limited by phonon-phonon, isotropic impurity and boundary scattering \cite{lindsay11}. 
The three-phonon scattering rates are incorporated in this method, as implemented in the the ShengBTE code \cite{ShengBTE}. Elaborate details on the work-flow of the three-phonon scattering rates can be found in reference \cite{ShengBTE} while Lindsay \cite{lindsay11} specifically discuses this for bulk h-BN.
The thermal conductivity matrix $\kappa_L^{\alpha \beta}$ is given as, 
\begin{eqnarray}\label{kl}
\kappa_L^{\alpha \beta}=\frac{1}{k_BT^2\Omega N}\sum_{s}f_0(f_0+1)(\hbar \omega_s)^2v_{s}^{\alpha} \tau_{s}^0 (v_s^{\beta}+\Delta_s^{\beta}).
\end{eqnarray}
$\kappa_L^{\alpha \beta}$ is then diagonalized to obtain the scalar lattice thermal conductivity $\kappa_L$ in a preferred direction in the $xy$ plane.
In Eq. \ref{kl} $\Omega$ is the volume of the unit cell, $N$ denotes the number of $q$-points in the Brillouin zone sampling. $f_0 = {1 / (e^{\hbar \omega_s/k_B T} - 1)}$ is the Bose-Einstein distribution function, $\tau_s^0$ is the relaxation time for the mode $s$ with phonon frequency $\omega_s$, $v_s$ is the phonon group velocity, and $\Delta_s$ denotes the measure of how much associated heat current deviates from the relaxation time approximation. Mathematically, $\Delta_s$ and $\tau_{\lambda}^0$ is expressed as \cite{ShengBTE},
\begin{eqnarray}
\Delta_{\lambda} &=& \frac{1}{N}\sum_{i=+,-}\sum_{\lambda^{'} \lambda^{''}} \Gamma^{i}_{\lambda \lambda^{'} \lambda^{''} } (\xi_{\lambda \lambda^{''}}F_{\lambda^{''}}-\xi_{\lambda \lambda^{'}}F_{\lambda^{'}}) \nonumber \\
&+& \frac{1}{N}\sum_{\lambda^{'}}\Gamma_{\lambda \lambda^{'}}\xi_{\lambda \lambda^{'}}F_{\lambda^{'}} \\
\frac{1}{\tau_{\lambda}^0} &=& \frac{1}{N}(\sum_{\lambda^{'} \lambda^{''}}^{+}\Gamma_{\lambda \lambda^{'} \lambda^{''}}^{+} + \frac{1}{2}\sum_{\lambda^{'} \lambda^{''}}^{-}\Gamma_{\lambda \lambda^{'} \lambda^{''}}^{-} + \sum_{\lambda^{'}}\Gamma_{\lambda \lambda^{'}})
\end{eqnarray}
here $\lambda$($\lambda^{'}$,$\lambda^{''}$) represents the phonon branch index $s$($s^{'}$,$s^{''}$) and wave vector $q$($q^{'}$,$q^{''}$) while $\xi_{\lambda \lambda^{'}}$ and $F_{\lambda}$ is short-hand for $\frac{\omega_{\lambda^{'}}}{\omega_{\lambda}}$ and $\tau_{s}^0 (v_s^{\beta}+\Delta_s^{\beta})$ respectively. 
The three-phonon scattering rates denoted by $\Gamma^{i}_{\lambda \lambda^{'} \lambda^{''} }$($i = +,-$) and the scattering probabilities due to isotopic disorder denoted by $\Gamma_{\lambda \lambda^{'}}$ have the following expressions,
\begin{eqnarray}\hspace{-2em}
\Gamma^{\pm}_{\lambda \lambda^{'} \lambda^{''}} &=& \frac{\hbar \pi}{4 \omega_\lambda \omega_{\lambda^{'}}\omega_{\lambda^{'}}} \Big[\substack{f_0(\omega_{\lambda^{'}})-f_0(\omega_{\lambda^{''}}) \label{ae} \\ f_0(\omega_{\lambda^{'}})+f_0(\omega_{\lambda^{''}}+1)}\Big]  \nonumber \\
 &\times& \big|V_{\lambda \lambda^{'} \lambda^{''}}\big|^2\delta(\omega_\lambda \pm \omega_{\lambda^{'}} + \omega_{\lambda^{''}}) \\
 \Gamma_{\lambda \lambda^{'}} &=& \frac{\pi \omega^2}{2}\sum_{i}f_s(i)\bigg[1-\frac{M_s(i)}{\overline{M}(i)}\bigg]^2 \nonumber \\ &\times& \big|e^{*}_{\lambda}\cdot e_{\lambda}\big|^2 \delta (\omega_{\lambda} - \omega_{\lambda^{'}}).
\end{eqnarray}
Where $V^{\pm}$ is the scattering matrix element and is expressed in terms of the anharmonic IFCs ($\Phi$), eigen functions ($e$) and mass ($M$) of an atom as
\begin{eqnarray}
V_{\lambda \lambda^{'} \lambda^{''}} = \sum_{i,j,k}\sum_{\alpha \beta \gamma} \frac{\Phi_{ijk}^{\alpha \beta \gamma}e_{\lambda}^{\alpha}e_{\lambda^{'}}^{\beta}e_{\lambda^{''}}^{\gamma}}{\sqrt{M_i M_j M_k}}.
\end{eqnarray}
In the above expression, $i,j,k$ run over the atomic indices and $\alpha, \beta, \gamma$ are the Cartesian coordinates.
$\overline{M} = \sum_s f_s(i) M_s(i)$ is the average of masses ($M_s(i)$) of isotopes $s$ of the atoms $i$ having
 a relative frequency $f_s$.
$\Gamma^{+(-)}$ represents the absorption (emission) processes. A phonon which is a result of the absorption
process is a combined energy of two incident phonons, {\it i.e.} $\omega_{\lambda} + \omega_{\lambda^{'}} = \omega_{\lambda^{''}}$. Similarly, the emission process depicts the energy of an incident phonon being separated 
among two phonons, $\omega_{\lambda} = \omega_{\lambda^{'}} + \omega_{\lambda^{''}}$. Therefore in eq.\ref{ae} it is easy to see that the Dirac delta function, $\delta(\omega_\lambda \pm \omega_{\lambda^{'}} + \omega_{\lambda^{''}})$, imposes the conservation of energy in the absorption and emission processes.

It should be noted that the relaxation times is calculated in the ShengBTE code 
using an iterative approach by solving the phononBTE starting with the zeroth-order approximation, $\Delta_{\lambda} = 0$, also known as the RTA solution. These iterations continue till two successive values of $\kappa_L$ differ by $10^{-5}$ Wm$^{-1}$K$^{-1}$.
The interatomic third-order force constants (IFCs) are calculated using a real space supercell approach.

Length dependent thermal conductivity is then calculated by taking into account only phonons with a mean free path (MFP) below a certain threshold value.
This is done by calculating the cumulative lattice thermal conductivity with respect to the allowed MFP. 
Furthermore, there have been recent advanced experimental techniques proposed to measure the cumulative $\kappa_L$ as a function of phonon mean free path \cite{minnich11,regner13,johnson13}.

In order to compare our calculations to the lengths corresponding to experimental measurements, we fit the cumulative thermal conductivity in the form \cite{ShengBTE},
\begin{eqnarray} \label{cum-k}
\kappa_L(L) = \frac{\kappa_{L_{max}}}{1+\frac{L_0}{L}}.
\end{eqnarray}
where $L_0$ is a fitting parameter. $\kappa_L$ corresponding to a given length is calculated over a temperature range using Eq. \ref{cum-k} and the thermodynamic limit of the thermal conductivity ($\kappa_{L_{max}}$) is the value of $\kappa_L$ as $L \rightarrow \infty$.

\subsubsection{Callaway-Klemens approach (Analytical and numerical solutions)}
In the Callaway-Klemens's \cite{callaway59,klemens58} approach which has been modified by Nika {\it el al} \cite{nika2009}, 
the expression for thermal conductivity along $x$ and $y$ directions for two-dimensional layered materials, 
according to the relaxation time approximation (RTA) to BTE and isotropic approximation to phonon dispersion is given by,
\begin{eqnarray} \label{k}
\kappa &=& {1\over 4\pi k_B T^2 N \delta}  \nonumber \\
 &\times& {\sum\limits_{s} \int\limits_{q_{min}}^{q_{max}}[\hbar \omega_s(q)]^2 v_s^2(q) \tau_{U,s}(q)\frac{e^{\frac{\hbar \omega_s(q)}{k_B T}}}{[e^{\frac{\hbar \omega_s(q)}{k_B T}}-1]^2} q dq},
\end{eqnarray}
where $k_B$ is the Boltzmann constant, $\hbar$ is the reduced Planck constant, $T$ is the absolute temperature, $N$ is the number of layers, $\delta$ is the distance between two consecutive layers, $\omega_s (q)$ and $v_s(q)$ are the phonon frequency and velocity corresponding to the branch $s$ at phonon wave vector $q$. 
The wave vector corresponding to the Debye frequency and low cut-off frequency are denoted by $q_{max}$ and $q_{min}$, respectively. The method to calculate the low cut-off frequency will be discussed shortly.
$\tau_{U,s}$ is the three-phonon Umklapp scattering corresponding to branch $s$ at the wave vector $q$ expressed as,
\begin{eqnarray}\label{tau}
\tau_{U,s} = \frac{Mv_s^2(q)\omega_{D,s}}{\gamma_s^2(q) k_B T \omega_s(q)^2}.
\end{eqnarray}
Here, $M$ is total mass of the atoms in the unit cell, $\gamma_s(q)$ is the mode and wave vector dependent Gr\"{u}neisen parameter.

The validity of the form of relaxation time in the Umklapp scattering in eq. \ref{tau} for a 2D and 3D material was originally proposed by Klemens {\it et. al.}\cite{klemens94}, where phonons were treated by a two-dimensional Debye model.
This sets up a mode for the thermal conductivity in terms of a 2D phonon gas. 
On the basis of the phonon frequency dependence of the specific heat and mean free path, the form of $\tau_{U,s}$ in eq. \ref{tau} is valid for both 2D and 3D. 
Moreover, the calculations by Shen {\it et. al.} \cite{shen14} use the same form to describe the relaxation time of the Umklapp process for graphene and their results, when $\tau_{U,s}$ is multiplied by a factor of 3, are consistent with the paper of Lindsay {\it et. al.} \cite{lindsay11} which solves the phonon BTE beyond the RTA. Since eq. \ref{tau} cannot determine whether the U-processes are forbidden or not, the factor of 3 is added due to the symmetries seen in graphene which is explained in detail later.

Gr\"{u}neissen parameter ($\gamma_s(q)$) and the Debye frequency ($\omega_{D,s}$) corresponding to the branch $s$ 
is calculated by solving,
\begin{eqnarray}\label{wD}
\frac{A}{2\pi}\int\limits_0^{\omega_{D,s}}q\Big|\frac{dq}{d\omega}\Big|d\omega = 1,
\end{eqnarray}
where $A$ is the area of the unit cell.

The acoustic branches for in-plane modes for SLBN, BLBN, 5LBN and Bulk-{\it h}BN are linear whereas the out-of-plane acoustic mode have a quadratic behavior and hence for a simplified analytical solution we express the phonon frequencies as
\begin{eqnarray}
\omega_s(q) &=& v_s q \Rightarrow [s = {\rm LA, TA}] \label{wLATA} \\
&=& \alpha q^2 \Rightarrow [s = {\rm ZA}] \label{wZA}
\end{eqnarray}
Substituting these values in Eq. \ref{wD}, we find the Debye frequency is given by
\begin{eqnarray}\
\omega_{D,s} &=& 2v_s\sqrt{\frac{\pi}{A}} \Rightarrow [s = {\rm LA, TA}] \\
&=& \frac{4\pi \alpha}{A} \Rightarrow [s = {\rm ZA}]
\end{eqnarray}
The mode dependent anharmonic (Gr\"{u}neissen) parameters were calculated by applying a biaxial strain of $\pm$ 0.5\% to each of the structures.
Fig. \ref{gp} shows that the Gr\"{u}neisen parameter for the in-plane modes have a slight deviation from its 
average value along the $\Gamma$ to K direction.
Therefore assuming a constant value for $\gamma_s$ ($s$=LA,TA), Nika {\it et al} \cite{nika2009} have derived the
following analytical solution for $\kappa$ associated with a particular mode $s$.
\begin{eqnarray} \label{LATA}
\kappa_s=\frac{M\omega_{D,s} v_s^2}{4\pi T (N\delta) \gamma_s^2}\ [{\rm ln}(e^x-1)+\frac{x}{1-e^x}-x]\Bigg|^\frac{\hbar \omega_{D,s}}{k_B T}_{\frac{\hbar \omega_{min,s}}{k_B T}}
\end{eqnarray}
Since there is no ZO$'$ branch in SLBN, the low bound cut-off frequency cannot be introduced in analogy to that of bulk graphite.
One can however avoid the logarithmic divergence by restricting the phonon mean free path on the boundaries of the sheets \cite{nika09}.
This is accomplished by selecting the mode dependent low cut-off frequency ($\omega_{s,min}$) from the condition that the mean free path cannot be greater in size than physical length $L$ of the sheet, {\it i.e,}
\begin{eqnarray} \label{wmin}
\omega_{s,min}=\frac{v_s}{\gamma_s}\sqrt{\frac{Mv_s\omega_{D,s}}{k_B T L}}
\end{eqnarray}

In the spirit of in-plane thermal conductivity study we extend our calculations to find an analytical form to 
the flexural phonons modes since the contribution from these branches are vital to the total thermal conductivity.
Unlike for the case of in-plane modes, the Gr\"{u}neisen parameters for the acoustic out-of-plane ZA modes have a strong $q$-dependence. From Fig. \ref{gp} it can be seen that the expression 
\begin{eqnarray} \label{gZA}
\gamma_{ZA}=\frac{\beta}{q^2},
\end{eqnarray}
is a very good fit to the actual wave vector dependent Gr\"{u}neisen parameters. Substituting eq. \ref{gZA} and eq. \ref{wZA} into eq. \ref{k} and making a transformation, $x=\frac{\hbar \omega}{k_B T}$, the analytical form for $\kappa_{ZA}$ is given by
\begin{eqnarray} \label{ZA}
\kappa_{ZA} &=& \frac{2M\omega_D k_B^3T^2}{\pi N \delta \beta^2 \hbar^3 \alpha} \int\limits_0^{\frac{\hbar \omega_D}{k_B T}}x^4 \frac{e^x}{[e^x-1]^2}dx \nonumber \\
&=& \frac{2M\omega_D k_B^3T^2}{\pi N \delta \beta^2 \hbar^2 \alpha}\; G\Big(\frac{\hbar \omega_D}{k_BT}\Big), 
\end{eqnarray}
where the function $G(z)$ is expressed as
\begin{eqnarray}
G(z) &=& \frac{-4\pi^4}{15} + \frac{e^z z^4}{1-e^z} + 4z^3{\rm ln}(1-e^z) \nonumber \\
  &+& 12z[z {\rm Li}_2(e^z)- 2{\rm Li}_3(e^z)] + 24{\rm Li}_4(e^z).
\end{eqnarray}
Here, the polylogarithm function is defined as, ${\rm Li}_n(z)=\sum\limits_{i=1}^{\infty} \frac{z^i}{i^n}$.


\section{Results and discussions}
\subsection{Phonon dispersion and density of states}
Accurate calculations of the harmonic second order IFCs are necessary for a precise description and understanding of the thermal conductivity. 
Deviations due to numerical artifacts from the expected behavior of acoustic modes can lead to incorrect results especially for 2D marterials \cite{jesus16}. The full structural relaxation of SLBN, BLBN, 5LBN and Bulk-{\it h}BN yield a lattice constant ($a_0$) of 2.49 \AA. The interlayer spacing ($c$) for MLBN is found to be  3.33 \AA. The experimentally measured $a_0$ is 2.50\ \AA\ \cite{kern99} and the ratio of interlayer spacing and the lattice constant ($\frac{a_0}{c}$) is 1.332 \cite{kern99} which is in excellent agreement with our calculated value of 1.337.

\begin{figure}[!htbp]
\centering \includegraphics[scale=0.36]{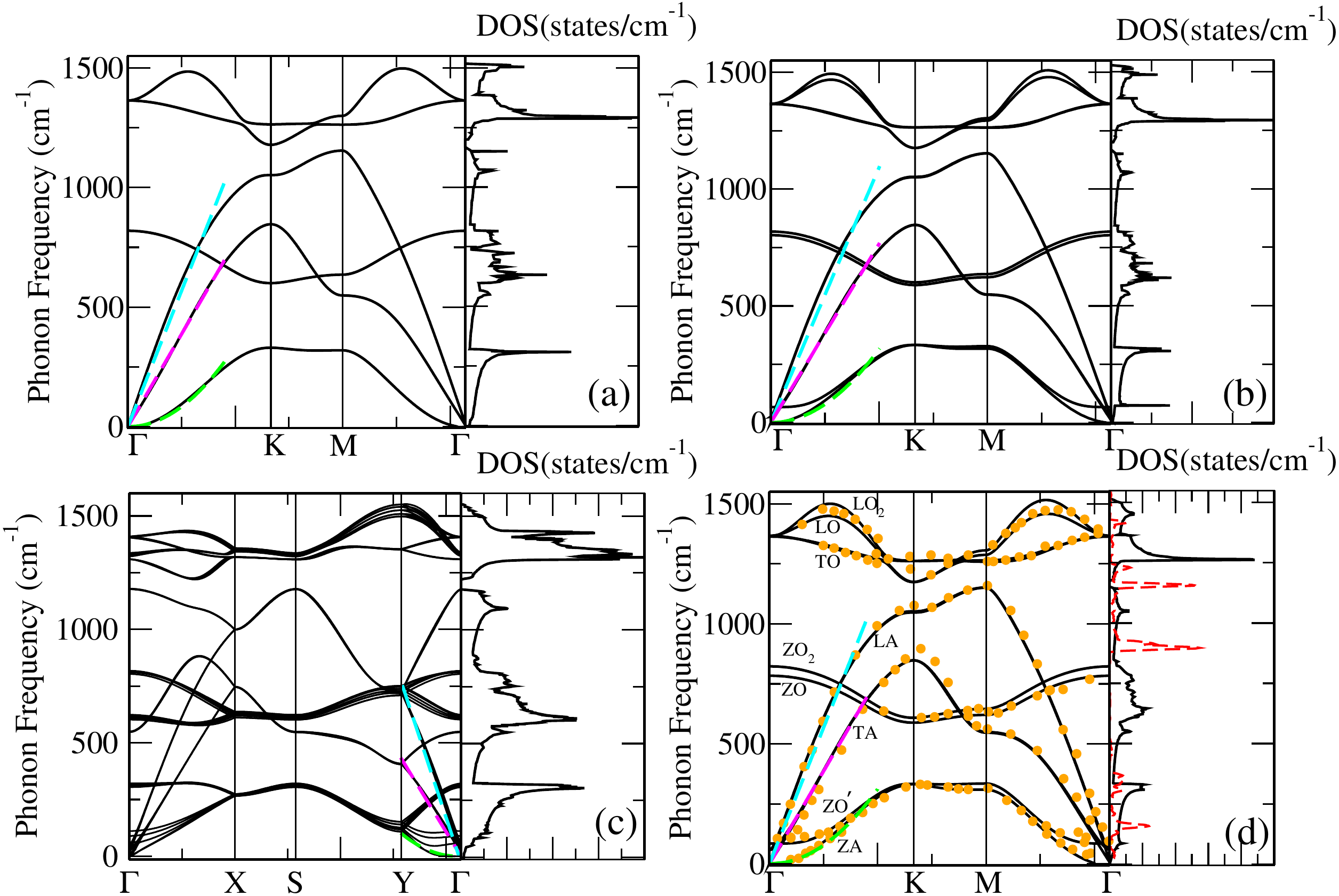}
\caption{\label{phdos} The calculated phonon dispersion (left) and phonon density of states (right) of (a) SLBN, (b) BLBN, (c) 5LBN and (d) Bulk-{\it h}BN along with experimental data (orange circles) \cite{serrano07}. 
The phonon dispersion were calculated along the high-symmetry points of the 2D Brillouin zone ($q_z = 0$) corresponding to the hexagonal cell for SLBN, BLBN, and Bulk-{\it h}BN and orthorhombic cell for 5LBN.
We also plot in (d) the two-phonon DOS shown for Bulk-{\it h}BN in red dashed line. The cyan, magenta and green curves in (a,b,c,d) are the best linear and quadratic fit to the phonon dispersion referring to LA, TA and ZA modes, respectively.}
\end{figure}

The calculated phonon dispersion and phonon density of states are shown
 in Fig. \ref{phdos} for (a) SLBN, (b) BLBN, (c) 5LBN and (d) Bulk-{\it h}BN along the high symmetric $q$-points in the irreducible hexagonal and orthogonal Brillouin zone (BZ)  together with some available experimental data for Bulk-{\it h}BN \cite{serrano07}. As usually seen for acoustic modes, the in-plane longitudinal (LA) and transverse (TA) modes show a linear $q$ dependence at the long-wavelength limit while the out-of-plane (ZA) mode shows a quadratic ($q^2$) dependence. This quadratic dependence, which is a typical feature of layered crystals, is due the rotational symmetries of the out-of-plane phonon modes. 

For SLBN, there are six modes for each wave vector, three acoustic (LA,TA,ZA) and three optical (LO,TO,ZO). At the $\Gamma$ point the optical LO and TO modes are degenerate.
For BLBN, if the two SLBN layers are far apart, effects due to their interlayer coupling can be neglected and the phonon dispersion will be exactly as what is seen in SLBN. However, when these two SLBN come closer,
due to the interlayer coupling, the two-fold degeneracy is removed giving rise to in-plane and out-of-plane phase modes. The LA and TA modes are not perturbed much implying that the main effect of the interlayer interactions is due to the ZA modes. This is because the transverse motion of atoms in both the layers associated with these modes interact strongly with each other.
The same reasons hold on why 5LBN has one zero and four raised frequencies at the $\Gamma$ point.
In Bulk-{\it h}BN, there are four atoms per unit cell and the two atoms in each layer are now inequivalent therefore doubling each of the acoustic and optical modes. The acoustic modes at the zone boundaries fold back to the zone centre as two rigid layer modes \cite{tan12}, {\it viz}, an optically Raman inactive and an Raman active mode. The Raman active LA$_2$ and TA$_2$ modes are doubly degenerate at the $\Gamma$ point having a finite value mentioned in Table \ref{pd}.
The layered breathing modes for MLBN are denoted by ZO$^{'}$ for BLBN and Bulk-{\it h}BN and ZO$^{'}_{i}$ ($i=1,2,3,4$) for 5LBN.

The symmetries of SLBN, BLBN, 5LBN and Bulk-{\it h}BN structures at $\Gamma$ can be described using the character table shown in table \ref{ct}.
Using a standard group theoretical technique  (see Appendix), it can be shown that for 
Bulk-{\it h}BN and BLBN that the 12 phonon modes are decomposed into the following irreducible representations: 2(A$_{\rm 2u}$ + B$_{\rm 1g}$ + E$_{\rm 2g}$ + E$_{\rm 1u}$) and 2(A$_{\rm 2u}$ + E$_{\rm g}$ + A$_{\rm 1g}$ + E$_{\rm u}$). Similarly for SLBN, the irreducible representation is A$_{\rm 2u}$ + B$_{\rm 1g}$ + E$_{\rm 2g}$ + E$_{\rm 1u}$ for the six phonon modes and 5LBN has an irreducible presentation given by 4(A$_1^{'}$ + E$^{''}$) +  6(A$_2^{''}$ + E$^{'}$).
Transitions corresponding to the basis $x,y,z$ ($xy,yz,z^2$, etc.) are Infrared (Raman) active. 
Those that are neither Infrared or Raman are the silent modes.
Due to the momentum conservation requirement ($q=0$), the first-order Raman scattering process is limited to the phonons at the center of the Brillouin zone. We therefore compare our calculated frequencies at the $\Gamma$ point  corresponding to A$_{\rm 2u}$, E$_{\rm 1u}$, A$_{\rm 2}^{''}$, E$^{'}$, and E$_{\rm u}$  to the infrared experimental data and E$_{\rm 2g}$, E$^{''}$, A$_{\rm 1g}$, E$_{\rm g}$, and A$_{\rm 1}^{''}$ to the Raman experimental data as shown in table \ref{pd}.

\begin{table*}[t] 
\caption{\label{pd} Experimentally measured Raman and Infrared phonon frequencies for bulk-{\it h}BN and those obtained from present calculations for all the systems studied are shown at the $\Gamma$ point in the BZ. Previously calculated values for bulk-{\it h}BN are also shown together with the experimental data for comparison.}
\begin{tabular}{lcccc}
\hline\hline
Mode & \begin{tabular}{c}Expt. (\& Prev. calculated$^{\rm a}$ )\\ $\omega$ (cm$^{-1}$) \end{tabular} &\begin{tabular}{c}Bulk-{\it h}BN\\ (Sym.) \end{tabular} & \begin{tabular}{c} BLBN\\ (Sym.) \end{tabular}& \begin{tabular}{c}SLBN\\ (Sym.) \end{tabular} \\
\hline
LA$_2$ \& TA$_2$ & 51.62$^{\rm b}$ (52.43) & 58.55  (E$_{{\rm 2g}}$) & 25.73 (E$_{{\rm g}}$) & - \\
ZO$'$  & Silent (120.98) & 85.01 (B$_{{\rm 1g}}$) & 66.54 (A$_{{\rm 1g}}$) & - \\
ZO  & 783.16$^{\rm c}$ (746.87) & 784.05 (A$_{{\rm 2u}}$)  & 803.01 (A$_{{\rm 2u}}$) & 819.37 (A$_{{\rm 2u}}$) \\
ZO$_2$  & Silent (809.78) & 823.17 (B$_{{\rm 1g}}$) & 818.25 (A$_{{\rm 1g}}$) & - \\
LO  & 1366.30$^{\rm b}$, 1370.33$^{\rm c}$, 1363.88$^{\rm d}$ (1379.20) & 1363.80 (E$_{{\rm 2g}}$) & 1364.45 (E$_{{\rm g}}$) & 1363.88 (E$_{{\rm 2g}}$) \\
TO & 1367.10$^{\rm c}$ (1378.4) & 1366.95  (E$_{{\rm 1u}}$) & 1365.66 (E$_{{\rm u}}$) & 1363.88 (E$_{{\rm 1u}}$) \\
\hline
\  & \begin{tabular}{c} LA \& TA (cm$^{-1}$)\\ (Point Group Symmetry) \end{tabular} & \begin{tabular}{c} LO (cm$^{-1}$)\\ (P.G. Symmetry) \end{tabular} & \begin{tabular}{c} TO (cm$^{-1}$)\\ (P.G. Symmetry) \end{tabular}  & \begin{tabular}{c} ZO (cm$^{-1}$)\\ (P.G. Symmetry) \end{tabular} \\
\hline
\ &  14.60 (E$^{''}$) & 1409.46 (E$^{'}$) & 1405.23 (E$^{'}$) & 817.59 (A$^{'}$) \\
\ &  31.10 (E$^{'}$) & 1408.91 (E$^{''}$) & 1404.94 (E$^{''}$) & 814.58 (A$^{''}$) \\
5LBN  & 38.95 (E$^{''}$) & 1408.71 (E$^{'}$) & 1404.81 (E$^{'}$) & 812.58 (A$^{'}$) \\
\ &  47.43 (E$^{'}$) & 1408.36 (E$^{''}$) & 1404.49 (E$^{''}$) & 810.27 (A$^{''}$) \\
\ & & 1405.57 (E$^{''}$) & 1404.40 (E$^{'}$) & 803.27 (A$^{'}$) \\
\hline\hline

\end{tabular}
\begin{tablenotes}
\item
$^{\rm a}$ From {\it ab initio} dispersion calculations, Ref. \cite{serrano07}. \\
$^{\rm b}$ Experimental Raman data, Ref. \cite{nemanich81}. \\
$^{\rm c}$ Experimental Raman and Infrared data, Ref. \cite{geick66}. \\
$^{\rm d}$ Experimental Raman data, Ref. \cite{reich05}.
\end{tablenotes}
\end{table*}

Raman spectroscopy is the most adaptable tool that offers a direct probe for multi-layered samples \cite{tan12}.
 Table \ref{pd} shows the transitions corresponding to the Inflared (E$^{'}$ and A$^{''}$) and Raman (E$^{''}$ and A$^{'}$) active modes in the case of 5LBN.
Further experiments for layered boron nitride would be required to verify the correctness of calculations. However, LDA with VdW interaction have shown to accurately describe the phonon dispersions for layered graphene when the geometry ({\it i.e.} interlayer distance) is represented correctly even though the local or semi-local exchange correlation functionals may not represent the interactions correctly \cite{tan12}. 

Another experimental technique to analyse the modes of a system is the second-order Raman spectroscopy in which the peaks are seen over the entire frequency range. 
Most of these peaks are in agreement with the phonon density of states when the frequency is scaled by a 
factor of $2\,$ \cite{serrano07,kern99}. We have hence plotted, to the right of our phonon dispersion, the frequency scaled DOS. However, as pointed out by Serrano {\it et al.}, peaks which are absent in the DOS can be seen in the second-order spectroscopy because the DOS does not take both overtones, {\it i.e.} summation of modes having the same frequencies, into account.
The two phonon density of states (DOS$_{2ph}$) are also essential for the understanding of phonon anharmonic decay \cite{cusco16}.
Experiments on the second-order Raman spectrum of h-BN has been performed by Reich {\it et al} \cite{reich05}. 
We show in Fig. \ref{phdos}(d) the two-phonon DOS \cite{esfarjani11},
\begin{eqnarray} \label{tpdos}
\hspace{-2em}{\rm DOS}_{2ph}(\omega) = \sum_{i,j} \delta(\omega - \omega_i - \omega_j) + \delta(\omega - \omega_i + \omega_j), 
\end{eqnarray}
for Bulk-{\it h}BN using our calculated harmonic interactions.
The peaks seen experimentally \cite{reich05} at 1639.4 cm$^{-1}$, 1809.907 cm$^{-1}$ and 2289.8068 cm$^{-1}$
are absent in the DOS. However, these large spectral features are now observed at 1680.4 cm$^{-1}$, 1821.2 cm$^{-1}$ and 2306.7 cm$^{-1}$, due to two phonon DOS (DOS$_{2ph}$).

\subsection{Thermal conductivity calculated using real space supercell approach}
\begin{figure}[b]
\centering \includegraphics[scale=0.33]{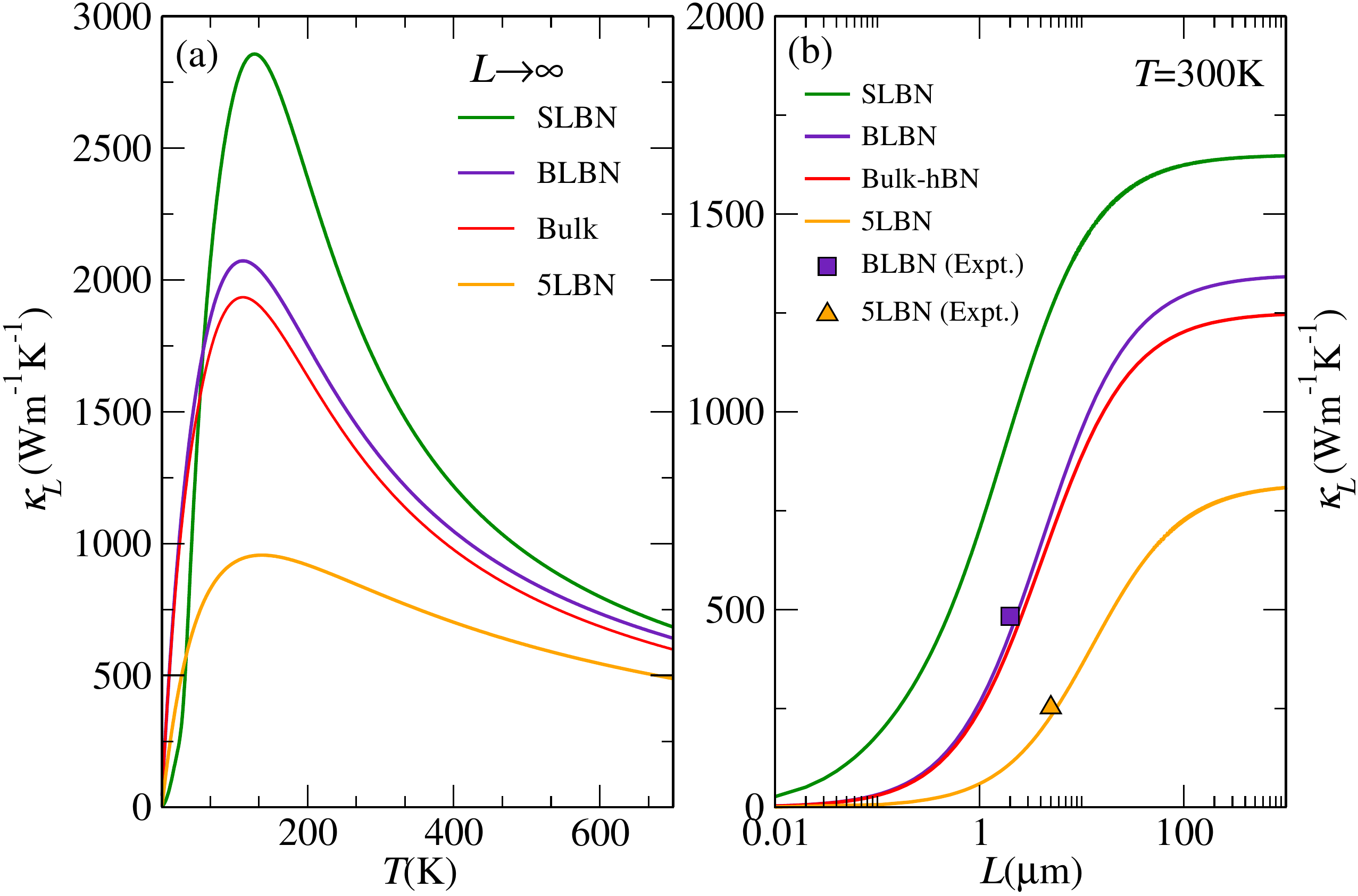}
\caption{\label{kfig} Calculated thermal conductivity of single and multilayer BN shown as a function of (a) temperature and (b) length, using the real space approach. In (a) the curves refer to the thermodynamic limit ($L\rightarrow \infty$). In (b) the sample length is in logarithmic scale. The square and triangle data points refer to experimental measurements for BLBN \cite{wang2016} and 5LBN \cite{jo13}, respectively.}
\end{figure}
In Fig. \ref{kfig} (a) and (b) we show the variation of thermal conductivity as a function of temperature ($T$) and
sample length, respectively.
The sample length is measured along the direction of the heat flow.
The theoretical computation was carried out using the interatomic force constants obtained from the real space approach and an iterative method in calculating the relaxation times as implemented in the ShengBTE code \cite{ShengBTE}. 
To have a broad understanding of the thermal conductivity, we study different types of possible unit cells, {\it i.e.},
MLBN considered here have even, odd and infinite number of layers since each unit cell has a different character table. Calculations were done using orthogonal cell for 5LBN  and hexagonal cells for SLBN, BLBN and bulk-{\it h}BN.
 The study was carried out over a wide range of sample lengths between 0.01 $\mu$m and 1000 $\mu$m with 0.1$\mu$m grid.
 The temperature of each sample was varied between 10 K to 1000 K with a grid of 10 K.
On plotting the thermodynamic limit ($L \rightarrow \infty$) for each of the system we find that $\kappa_L$ is practically independent of length for lengths greater than 100   $\mu$m.

Our recent results of $\kappa_L$ in the thermodynamic limit ($L\rightarrow \infty$) for monolayer and bilayer graphene \citep{RDSM17} are in excellent agreement with the recent experimental work of Li {\it et. al} \cite{hongyang2014},
whereas the thermodynamic limit for MLBN is much larger than some recent experimental measurements \cite{jo13,wang2016}.
Sample lengths used by Li {\it et. al.} were of the order of millimetres for the measurement of single and bilayer graphene
 while Jo {\it et. al.} and Wang {\it et. al.} have used sample lengths of 5 $\mu$m and 2 $\mu$m for 5LBN and BLBN, respectively.
As mentioned earlier, $\kappa_L$ does not vary much for lengths larger than 100 $\mu$m but is extremely sensitive when the lengths are between 1 and 100 $\mu$m.
Not surprising therefore, our thermodynamic limit of $\kappa_L$ are in good agreement for graphene but not for MLBN.

In order to compare our calculations to that of experiments, we calculate the cumulative lattice thermal conductivity at lengths corresponding to the sample lengths used in the experiments.
The cumulative $\kappa_L$ was calculated in the temperature range 10-1000 K. Fig. \ref{kfig} (b) shows the cumulative thermal conductivity at room temperature (RT).
\begin{figure}[b]
\centering \includegraphics[scale=0.33]{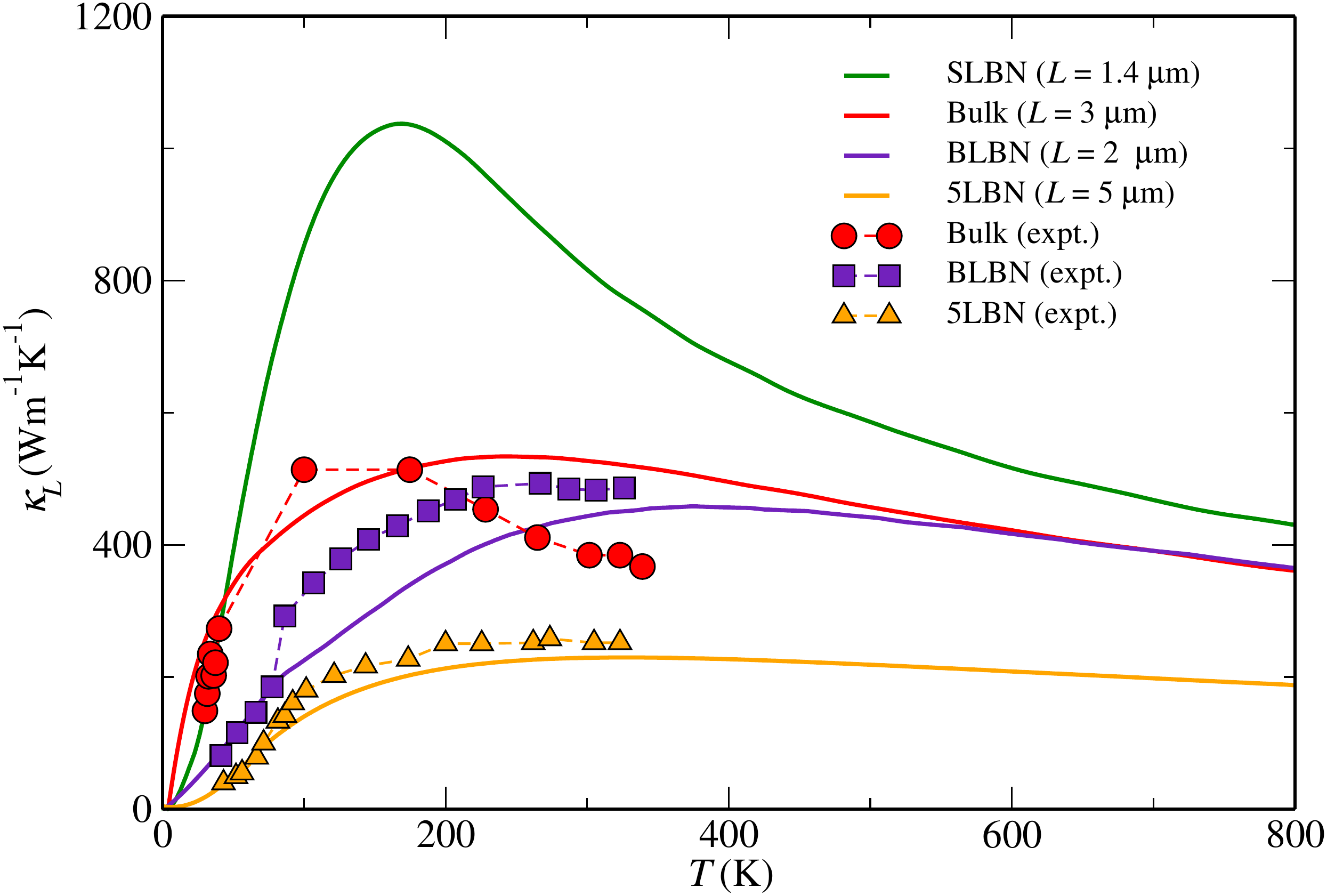}
\caption{\label{kLexpt} Calculated thermal conductivity of single and multilayer BN shown as a function of temperature at a constant length, using the real space approach. The square, circle, triangle data points refer to experimental measurements for BLBN \cite{wang2016}, Bulk {\it h}-BN \cite{sichel76} and 5LBN \cite{jo13}, respectively.}
\end{figure}

The curves in Fig. \ref{kLexpt}, are the calculated values of $\kappa_L$ at constant lengths which
are compared with the experimental observations \cite{wang2016,sichel76,jo13}.
For the lengths used in the experiments the magnitudes of $\kappa_L$ for Bulk-{\it h}BN and Bi-layer lie in between SLBN and 5LBN with SLBN (5LBN) being the highest(lowest). 
The maxima of $\kappa_L$ of $\sim$ 500 Wm$^{-1}$K$^{-1}$ for Bulk-{\it h}BN is found in the temperature range 250-300 K
 and tends to saturate to a value $\sim$ 450 Wm$^{-1}$K$^{-1}$. Experimentally \cite{sichel76} the maxima is found between 150-200 K and tends to saturate to a value $\sim$ 400 Wm$^{-1}$K$^{-1}$.
Lindsay {\it et. al.} \cite{lindsay11} varies the sample length and finds an excellent fit with the experimental data for $L=1.4 \mu$m. It must be noted that the sample length is not mentioned in the experimental reference 
\cite{sichel76} for Bulk-{\it h}BN.
As the length of the sample increases, the maxima of $\kappa_L(T)$ shifts towards the left, {\it i.e.} the maxima is found at a lower temperature.
Therefore for BLBN and 5LBN, where the lengths used in the experiments are larger than 1.4 $\mu$m, the maxima would be at lower temperatures, in total disagreement with the experiments \cite{wang2016,jo13}. 
Our calculations for BLBN and 5LBN are in excellent agreement with experiments for the same lengths.
Even though our calculated values diverge from the experimental measurements by Sichel {\it et. al.} \cite{sichel76} at higher temperatures, 
we believe that the behavior of $\kappa_L$ as calculated by us for bulk-{\it h}BN is correct.  
However, further experiments should throw more light on these discrepancies. 
It is our conjecture that $\kappa_L$ of Bulk-{\it h}BN should be similar to that of BLBN since the phonon dispersions 
in the two cases are very similar.

\begin{figure}[htbp!]
\centering \includegraphics[scale=0.33]{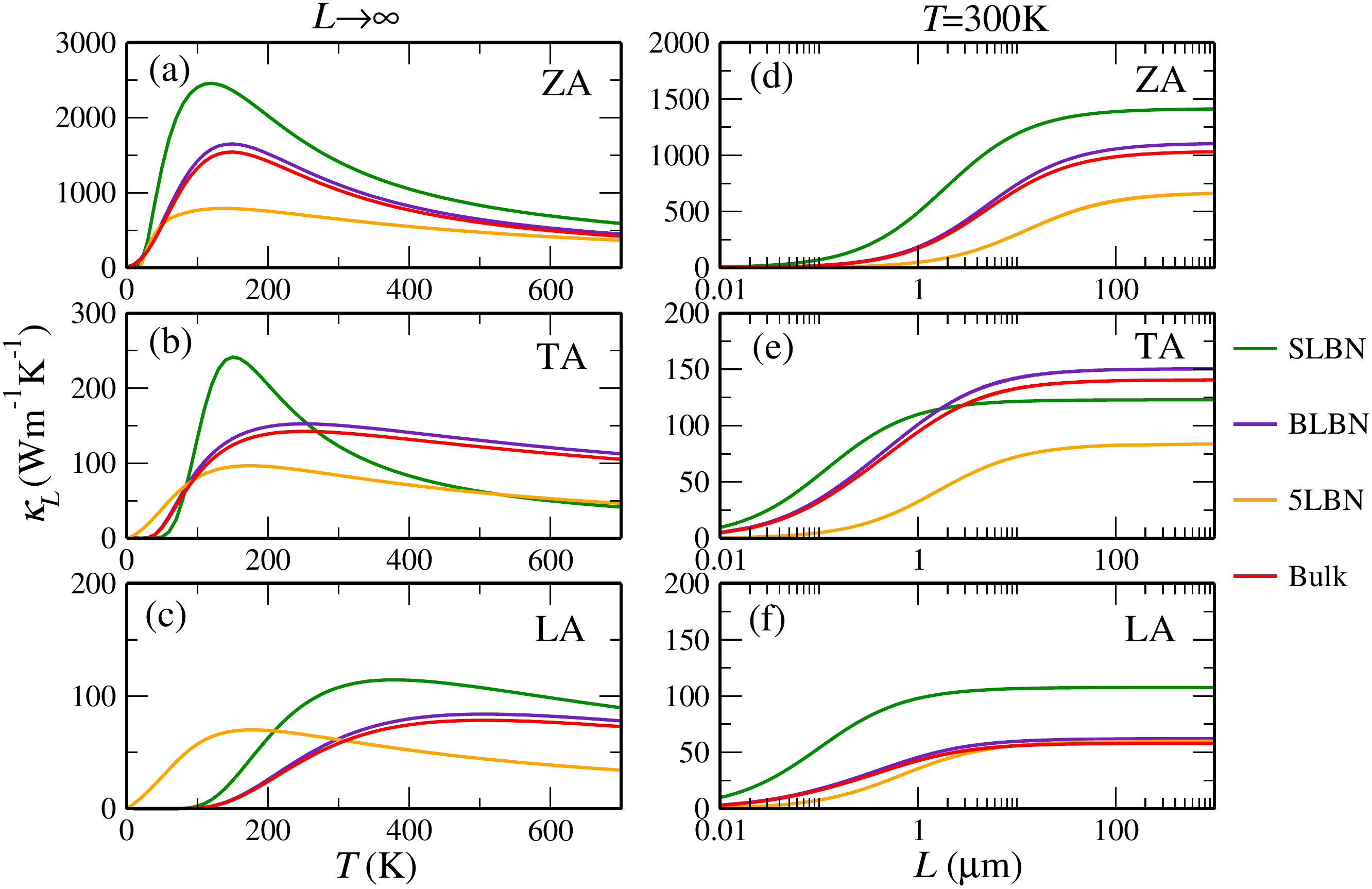}
\caption{\label{kLmode} Contribution to the thermal conductivity of single and multilayer BN from the acoustic modes; (a) ZA, (b) TA and (c) LA, shown as a function of temperature in thethermodynamic limit ($L\rightarrow \infty$), and as a function sample length at $T=300$K (d,e,f) using the real space approach.}
\end{figure}

In Fig. \ref{kLmode} we show the acoustic mode dependent contributions to the total thermal conductivity for SLBN and MLBN by solving the phonon BTE beyond the RTA. The out of plane mode is clearly seen to contribute the most to the lattice thermal conductivity for all the mentioned structures. 
For SLBN the contributions from the ZA, TA and LA modes to $\kappa_L$ at room temperature are $\sim$ 86.1 \%, 7.4 \% and 6.5 \%. A similar trend is observed in graphene \cite{lindsay2010}. Qualitatively one can understand why the ZA mode contributes the most to $\kappa_L$ by calculating the number of modes per frequency for each of the acoustic mode. 
Now the number of modes per frequency is proportional to the 2D density of phonon modes, $D_s(\omega) \propto \frac{q}{2\pi}\frac{dq}{d\omega}$, and hence the ratio of $D_{ZA}(\omega)$ and $D_{TA(LA)}(\omega)$
would give a measure of the contribution of the respective phonon modes.
Assuming a quadratic fit to the ZA dispersion, $\omega_{ZA}=\alpha q_{ZA}^2$, and a linear fit to the in-plane TA and LA phonon dispersion, $\omega_{TA(LA)}= v_{TA(LA)}q_{TA(LA)}$, the ratio of the density of phonon modes is $\frac{D_{ZA}}{D_{TA(LA)}} = \frac{v_{LA(TA)}^2}{2\alpha\omega_{LA(TA)}}$. Here $\alpha$ and $v_{LA(TA)}$ are fitting parameters to the phonon dispersions shown in Fig. \ref{phdos} and their values are shown in table \ref{para}. Substituting these values, it is evident that $\frac{D_{ZA}}{D_{TA(LA)}} \gg 1$ at the long wavelength limit suggesting that the major contributions to the 
lattice thermal conductivity are due to the out of plane modes.
Representing the ZA contribution of the thermal conductivity of MLBN at room temperature with respect to SLBN, we observe that $\kappa^{SLBN}_{ZA} = 1.28\kappa^{BLBN}_{ZA} = 2.17\kappa^{5LBN}_{ZA}$, suggesting that the significant decrease of $\kappa_L$ from SLBN to MLBN is because of the additional raised frequencies of the ZA layered breathing modes. 

Kong {\it et. al.} \cite{kong2009} reported that the lattice thermal conductivity of single layer graphene and bilayer are similar, $\kappa_L^{graphene} \approx \kappa_L^{bilayer}$, while Lindsay {\it et. al} \cite{lindsay2011} reported $\kappa_L^{graphene} \approx 1.37\kappa_L^{bilayer}$. 
The difference in their methodologies is that the latter has taken graphene symmetry into account, 
which is discussed in detail by Seol {\it et.al.} \cite{seol10} and Lindsay {\it et.al.} \cite{lindsay2011}. 
Besides the contribution due to the layer breathing out of plane modes, a decrease in $\kappa_L$ is also due to the violation of the selection rule \cite{seol10,lindsay2011} which is incorporated in the formalism in the super-cell real space approach. 
In Fig. \ref{kLmode} (d,e,f), we show the mode dependent $\kappa_L$ at room temperature as a function of sample length. At any given length, the maximum difference in $\kappa_L$ contributed from LA and TA modes for all 
the mentioned structure is $\sim$ 47 and 65 Wm$^{-1}$K$^{-1}$ respectively while that from the ZA mode is $\sim$ 750 Wm$^{-1}$K$^{-1}$, an order of magnitude larger, implying that the contribution from the in-plane thermal conductivity is almost independent of the number of layers. 
This characteristic has been seen using a Tersoff potential in the case of single and multilayered graphene and boron nitride \cite{lindsay2011, lindsay12}. 
This rapid decrease in $\kappa_L$ by increasing the number of layers, which is mainly due to the ZA mode, suggests that the interlayer interactions are short ranged, {\it i.e.}, the BN layers only interact with neighbouring BN layers \cite{lindsay2011}. 
In all of the structures, the contribution to $\kappa_L$ from the ZA mode have a stronger $L$ dependence as compared to the TA and LA modes, {\it i.e.}, the contributions from the in-plane modes saturate to their thermodynamic limit at a lower $L$ value as compared to the contributions from the out-of-plane modes. This is due to the larger intrinsic scattering times allowing the ZA phonons to travel ballistically and the relatively smaller scattering time which reflects the diffusive transport of the TA and LA phonons \cite{lindsay2011}.
Calculations based on the mode dependent contributions to $\kappa_L$ as a function of mean free path and recent advanced experimental techniques \cite{minnich11,regner13,johnson13} should motivate further studies in these directions. 

The in-plane phonon contributions having a small $L$ dependence in comparison to the contributions from the out of plane has been calculated for graphene recently using the Tersoff potential \cite{lindsay14} and their calculated cumulative mode dependent thermal conductivity behavior is in good agreement with our calculations for SLBN.

\subsection{Gr\"{u}neisen parameter}
\begin{figure}[t]
\centering \includegraphics[scale=0.32]{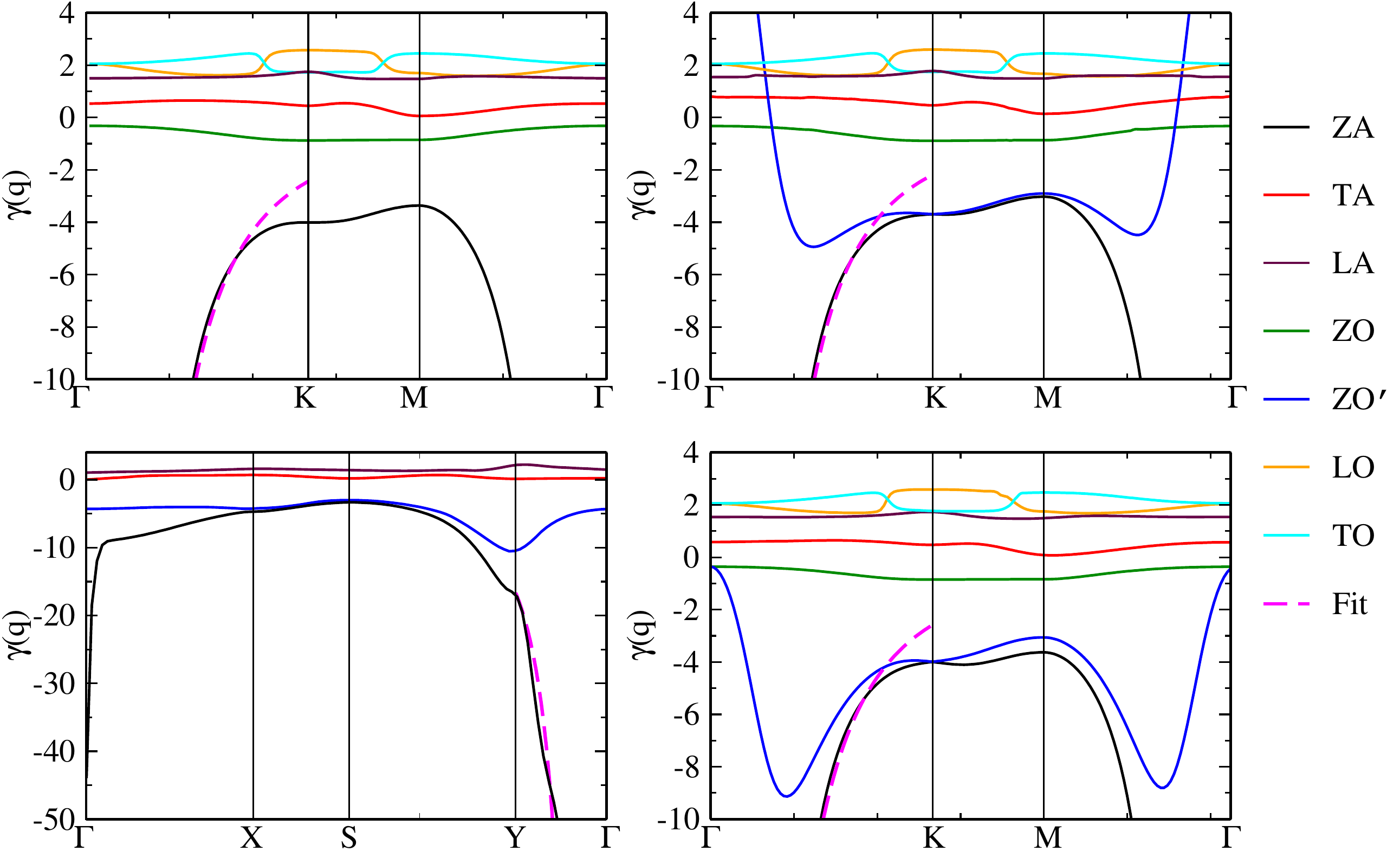}
\caption{\label{gp} Gr\"{u}enisen parameters of each mode for (a) SLBN, (b) BLBN, (c) 5LBN and (d) Bulk-{\it h}BN. The colour representation of each mode and fit are shown on the right. The magenta curves are the best fit to the ZA mode along the direction in the BZ chosen to calculate the lattice thermal conductivity.}
\end{figure}

Besides providing important information on the phonon relaxation time, the Gr\"{u}neisen parameter ($\gamma$) also provides information on the degree of phonon scattering and anharmonic interactions between lattice waves. Therefore, an accurate calculation of the lattice thermal conductivity ($\kappa_L$) would require a precise calculation of $\gamma$ since anharmonic lattice displacements play a vital role in calculations of $\kappa_L$. Fig. \ref{gp} displays the mode dependent $\gamma$ for SLBN, BLBN, 5LBN and Bulk-{\it h}BN along the high symmetric $q$ points. The anharmonic lattice displacements are carried out by dilating the unit cell by applying a biaxial strain of $\pm$ 0.5 \% and is expressed as,
\begin{eqnarray}\label{gamma}
\begin{split}
\gamma_s(q) & = \frac{-a_0}{2\,\omega_s(q)}\frac{\delta \omega_s(q)}{\delta a} \\
  & \approx \frac{-a_0}{2\,\omega_s(q)}
\Big[\frac{w_+ - w_-}{da}\Big]
\end{split}
\end{eqnarray}
This method has been used previously for graphite \cite{marzari05}, single and bi-layer graphene \cite{RDSM17} and MoS$_2$ \cite{cai10}.
where $\omega_s(q)$, $\omega_+$ and $\omega_-$ are the wave vector dependent phonon frequency of mode $s$, phonon frequency under positive and negative biaxial strain  respectively. $a_0$ and $d a$ are respectively the relaxed lattice constant and difference in lattice constants when under positive and negative biaxial strain.
We find that the acoustic modes correspond to the lowest Gr\"{u}neisen parameters which is in-line with experimentally measured $\gamma$ \cite{sanjurjo83}. 
As in the case of graphene, the out-of-plane acoustic transverse mode has the largest negative $\gamma$ parameters.

Positive (negative) Gr\"{u}neisen parameters suggest a decrease (increase) in phonon frequencies as the lattice
constant increases.
Near the long-wavelength limit, $\gamma_{ZO^{'}}$ for 5LBN is positive but becomes negative as we move along
the $\Gamma$ to Y direction in the BZ. $\gamma_{ZO^{'}}$, associating with the layer breathing mode suggests 
that due to the additional layers the atom vibrations along the perpendicular direction between them lose their
coherence and hence decreases the phonon frequencies when the system is under a biaxial strain.

As described in table \ref{ct}, E$_{\rm 2g}$, E$^{''}$, A$_{\rm 1g}$, E$_{\rm g}$, and A$_{\rm 1}^{''}$ are Raman active and hence in principle their Gr\"{u}neisen parameter can be calculated experimentally
using Raman spectroscopy.
There exist experimental data for bulk {\it h}-BN but to the best of our knowledge there does not exist experimental data for single or MLBN. We therefore compare our results to that of bulk-{\it h} BN. 

The lowest Gr\"{u}neisen parameters along the $\Gamma$-K-M directions for the TO and LO modes were found to be 1.72 and 1.59, respectively. 
Our calculations for these modes are only $\sim$ 1.1\% and $\sim$ 1.3\% larger than the experiment values of Sanjurjo {\it et al.} \cite{sanjurjo83} who have obtained the Gr\"{u}neisen parameters by measuring the pressure dependence of Raman lines.
The slight deviance from the experimental measured value could be because the measured values are for Zinc-blende-BN and not hexagonal BN.
\subsection{Analytical solutions to the Callaway-Klemens's Approach}
\begin{figure}[b]
\centering \includegraphics[scale=0.32]{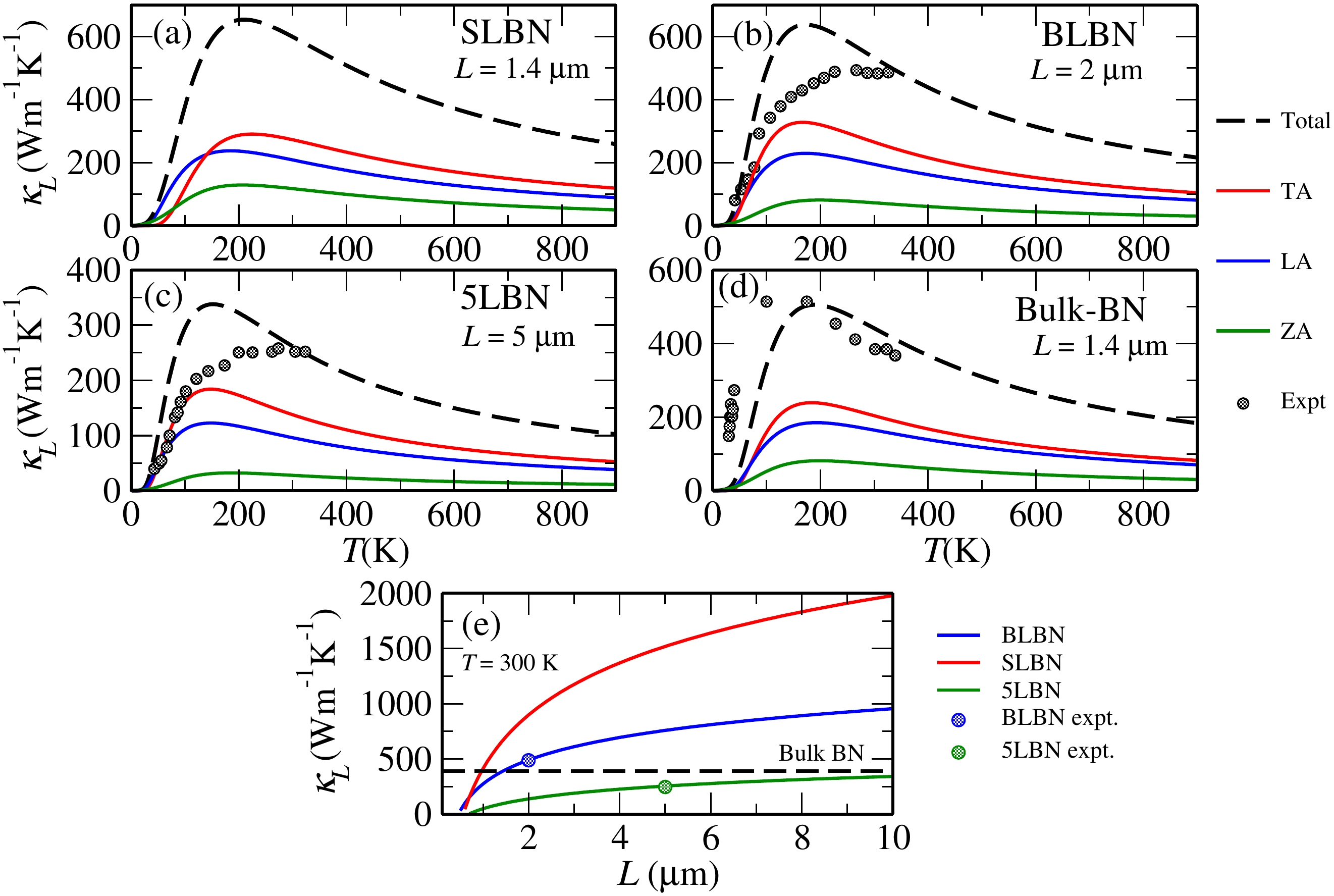}
\caption{\label{kaLATAZA} Acoustic modes and temperature dependence of lattice thermal conductivity for (a) SLBN (b) BLBN (c) 5LBN and (d) Bulk-{\it h}BN at a constant length. The theoretical calculations are carried out by using Eq. \ref{LATA} for the LA and TA modes while Eq. \ref{ZA} was used for the ZA mode.
The parameters used in our calculations are shown in Table \ref{para}. The colour representation for each mode are shown on the right. The black dots are the experimental measurements \cite{sichel76,jo13,wang2016}.
Length dependence is worked out by varying $L$ in Eq. \ref{wmin}}
\end{figure}

In order to compare the results obtained from the real space super cell approach (ShengBTE), we now study the mode, temperature and length dependence of single and MLBN calculated using the Callaway-Klemens's approach as described earlier.
We first obtain analytical solutions
 for each acoustic mode of the Phonon BTE by making some reasonable approximations to understand the basic behavior of temperature and length dependence of $\kappa_L$.
 In order to compare with the experimental results, we resort to exact numerical computation.
We have carried out all the length dependent calculations at a constant temperature for MLBN at RT.
The corresponding length dependent curves for MLBN are plotted in Fig. \ref{kaLATAZA} (e). 
 The parameters used in our study are shown in Table \ref{para}. 
\begin{table}[t]
\caption{\label{para} Relevant parameters used in the calculations for the analytical solutions of the lattice thermal conductivity.}
{\footnotesize \begin{tabular}{c|c|c|c|c|c|c}
\hline\hline
System & \begin{tabular}{c} $v_{LA}$ \\ (m/s)\end{tabular} & \begin{tabular}{c} $v_{TA}$ \\ (m/s)\end{tabular} & $\gamma_{LA}$ & $\gamma_{TA}$ &\begin{tabular}{c} $\alpha$  $\times$ 10$^{-7}$ \\ (m$^2$/s)\end{tabular} & \begin{tabular}{c} $\beta$ $\times$ 10$^{-20}$\\ (1/m$^2$)\end{tabular} \\
\hline
SLBN & 17020.1 & 11599.8 & 1.546 & 0.452 & 3.99  & -6.827  \\
BLBN & 16379.4 & 11474.9 & 1.585 & 0.5673 & 3.75  & -6.086  \\
5LBN & 21095 & 11420.6 & 1.48 & 0.424 & 4.2  & -6.348  \\
Bulk-{\it h}BN & 16379.4 & 11474.9 & 1.57 & 0.59 & 3.72  & -7.18  \\
\hline\hline
\end{tabular}}
\end{table}
Equations \ref{wLATA} and \ref{wZA} are plotted in Fig. \ref{phdos} and Equation \ref{gZA} is plotted in Fig. \ref{gp} to compare the analytical fit to the actual phonon dispersion and Gr\"{u}enisen parameters.

The individual contributions of each of the acoustic modes LA, TA, ZA and the sum of these, {\it i.e.} $\kappa_L$, for single and multilayered {\it h}-BN are shown in Fig. \ref{kaLATAZA} (a,b,c,d). 
The variation of $\kappa_L$ values for BLBN and Bulk-{\it h}BN are quite similar but are lower for 5LBN. This is in good agreement with experiments \cite{sichel76,jo13,wang2016}. 
In all cases it is observed that amongst the acoustic modes the TA contribution is the largest, ZA to be the least whereas LA contribution is somewhere in between.
It has been quite controversial as to which acoustic mode contributes the most to the total lattice thermal conductivity.
For example, while some reports \cite{kuang16,lindsay11,lindsay2011,seol10,lindsay2010} show that the contributions from ZA to be the most dominant, there are many other reports \cite{shen14,alofi13,kong2009,aksamija11,nika2009,nika11,nika12,wei14} that show exactly the opposite.
Our analytical results concur with the latter, {\it i.e.} the contribution from the ZA mode is the least.

The thermal conductivity for two-dimensional layered materials given by Eq. \ref{k} is derived assuming both phonon energy dispersions and phonon scattering rates are weakly dependent on the direction of the Brillouin zone \cite{nika2009}. The calculation of $\kappa_L$ should be independent of the direction chosen resulting in an isotropic in-plane scalar $\kappa_L$. 
Calculation of $\kappa_L$ should therefore be independent of direction chosen.
We move along the $\Gamma$ to K direction for systems in which a hexagonal unit cell is used and along $\Gamma$ to Y in the case of an orthorhombic unit cell. SLBN has the highest calculated $\kappa_L$, 5LBN has the least while $\kappa_L$ lies in between BLBN and Bulk-{\it h}BN. From Fig. \ref{kCK} it can be easily seen that for temperatures below 100K, the contribution to the total $\kappa_L$ is mainly due to the flexural ZA modes.

As in the case of graphene, SLBN can have a total of 12 processes involving the flexural phonons. 
However, Seol {\it el al} \cite{seol10} obtained a selection rule for the three-phonon scattering. This rule states that only an even number of ZA phonons is allowed to be involved in each process. Shen {\it et al} \cite{shen14} have listed four flexural allowed processes. Hence, the scattering rate of the Umklapp phonon-phonon process for the acoustic flexural branch is multiplied by a factor of $\frac{4}{12}$ and the relaxation time for the ZA mode becomes 3 times of that of Eq. \ref{tau}. Therefore besides having a larger velocity and a smaller averaged Gr\"{u}neisen parameters compared to the other systems, the major contribution for an increased $\kappa_L$ is due to the symmetry of the ZA mode. 

Phonon dispersions and Gr\"{u}neisen parameters for BLBN and Bulk-{\it h}BN are very similar which explains why their calculated $\kappa_L$ have the same magnitude. In the case of 5LBN, there are additional five low frequency modes (also termed as layer-breading modes), which arise due to the interlayer moment. Due to this change in phonon dispersion, more phase-space states become available for phonon scattering and therefore decreases $\kappa_L$ \cite{balandin11}.

\subsection{Numerical solutions to the Callaway-Klemens's Approach}
\begin{figure}[t]
\centering \includegraphics[scale=0.32]{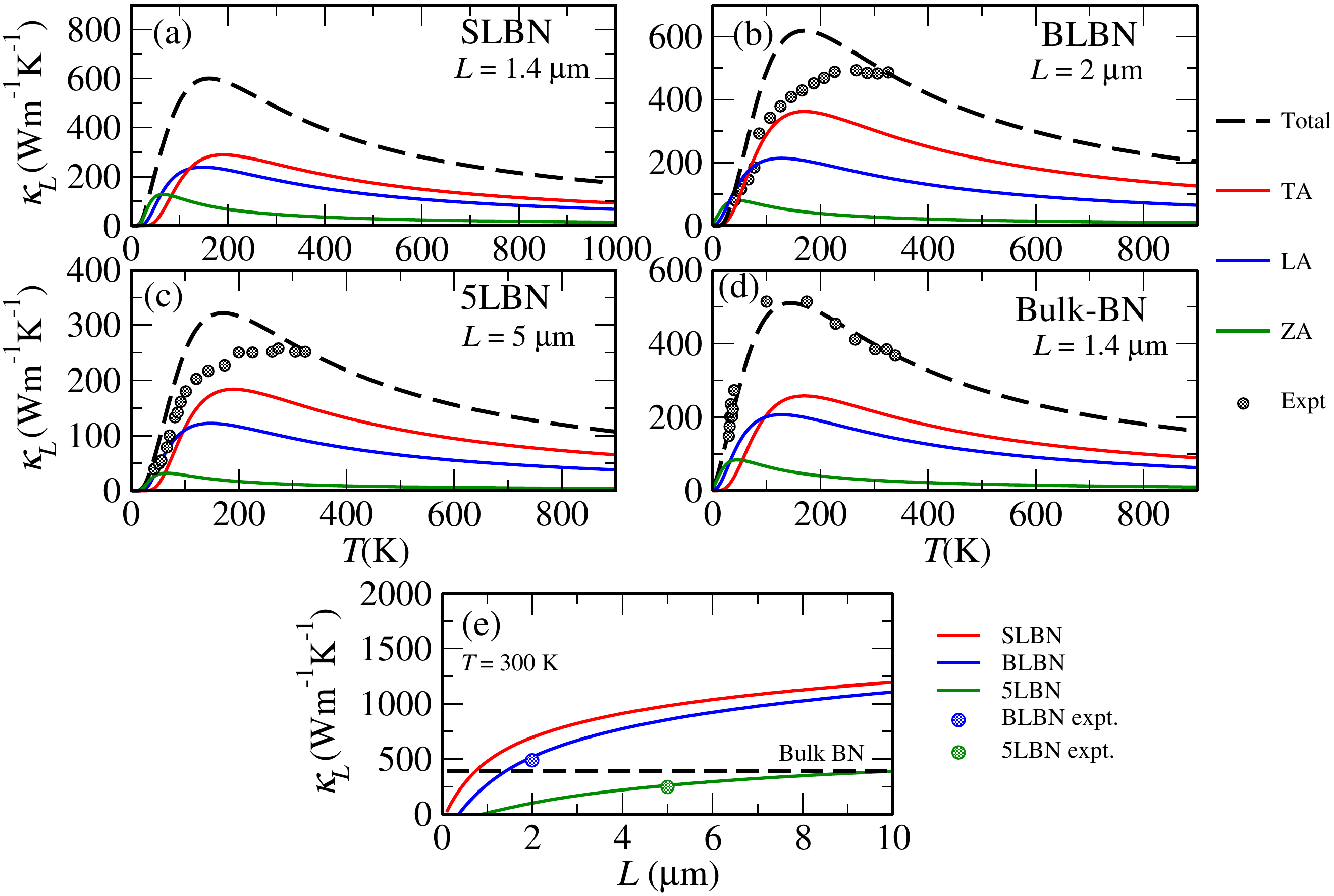}
\caption{\label{kCK} Acoustic modes and temperature dependence of lattice thermal conductivity for (a) SLBN (b) BLBN (c) 5LBN and (d) Bulk-{\it h}BN at a constant length. The theoretical calculations are carried out by solving Eq. \ref{k} numerically for each of the modes.
The colour representation for each mode are shown on the right. The black dots are the experimental measurements \cite{sichel76,jo13,wang2016}.
Length dependence is worked out by varying $L$ in Eq. \ref{wmin}.}
\end{figure}

Numerical calculations are carried out using the exact form of the phonon dispersion and Gr\"{u}neisen parameters as displayed in Fig. \ref{phdos} and Fig. \ref{gp} rather than the analytical form of the acoustic modes and averaged Gr\"{u}neisen parameters.
 We numerically solve Eq. \ref{k} for each of the modes at a constant sample length varying temperature as well as at a constant temperature varying lengths between 0.1 to 10 $\mu$m.
  These results are compared with experimental data \cite{sichel76,jo13,wang2016} and shown in Fig. \ref{kCK}.
Numerically calculated values of $\kappa_L$ are in better agreement with the experimental data as compared to the analytical form. We find the contribution from the ZA modes to dominate at lower temperatures but rapidly decreases as the temperature increases making the flexural modes contribute the least at relatively higher temperatures. 
This is in line with previous theoretical calculations \cite{aksamija11}.

\section{Summary}
Phonon dispersions using a LDA pseudopotential with vdW interactions, density of states (DOS), the Gr\"{u}neisen parameters and the lattice thermal conductivity have been calculated by the Callaway-Klemens and Real space super cell approach for SLBN, BLBN, 5LBN and Bulk-{\it h}BN.
Additionally, in the case of Bulk{\it h}-BN, we calculate the two-phonon DOS.
Irreducible representation using the character table at the $\Gamma$ point in the BZ for each of the systems have been derived in order to compare the symmetry modes with those obtained from Raman and infrared spectroscopy experiments.
Results from the investigations by EELS data, Raman, second-order Raman and Infrared spectroscopy are found to be in excellent agreement with the theoretical calculations based on the phonon dispersion, DOS and two-phonon DOS which rely on the harmonic second order inter atomic force constants.

Further, we have calculated the sample length and temperature dependence of lattice thermal conductivity 
by the real space super cell approach with the help of the second order IFCs calculated by DFPT.
Lattice thermal conductivity at the thermodynamic limit for each system has a maxima between the 110-150 K.
For sample sizes in the range 1-5 $\mu$m, $\kappa_L$ does not have a maxima. However with increase in temperature it tends to saturate at a value which is an order smaller than the
thermodynamic limit.

Our mode dependent calculations using the real space method suggests that the majority of the contribution to the thermal conductivity are due to the ZA phonons for all of the structures. The substantial decrease in $\kappa_L$ from single to MLBN is because of the additional layer breathing modes but mainly due to the fact that the interlayer interactions breaks the SLBN selection rule resulting in suppressing the ZA phonons contributions to $\kappa_L$ in MLBN.
Contribution to $\kappa_L$ from the in-plane modes are not sensitive to the number of layers and have a lower $L$ dependence compared to the out of plane modes.
This reduction in $\kappa_L$ from SLBN to MLBN which is mainly due to the ZA phonons indicate that the interlayer interactions are short ranged.
The $L$ dependence of the TA and LA contributions to $\kappa_L$ saturate to their thermodynamic limit faster than that of the contribution from the ZA phonons implying that the ZA phonons travel ballistically along the sample while the TA phonons travel diffusively.

Gr\"{u}neisen ($\gamma$) parameters were obtained from first principle calculation by applying a positive and negative biaxial strain.
For the in-plane acoustic modes, we find that $\gamma$ does not vary much from its mean value but the out-of-plane 
modes have a strong $q$-dependence.
Our calculated $\gamma$ values for Bulk-{\it h}BN at the $\Gamma$ point is $\sim$ 1\% larger than those obtained
from experiments which measures the pressure dependence of Raman lines.
$\gamma$ parameters for 5LBN suggest that due to the layer breathing modes, atoms along the perpendicular direction lose their coherence between each layer and decrease the phonon frequencies when under a
biaxial strain.

In comparison to the real space super cell approach, lattice thermal
conductivity has been calculated, both analytically and numerically, using Callaway-Klemens formalism.
To obtain analytical solution of the phonon, we make a linear fit to the LA and TA modes,
a quadratic fit to the ZA mode, and use an averaged value for its Gr\"{u}neisen parameters for the $\gamma$ parameters corresponding to the in-plane acoustic modes and an inverse square wave-vector dependence $\gamma$ for the out-of-plane modes. 
Theoretical results for sample length and temperature dependence of $\kappa_L$ are in good agreement with
experimental observation.
The phonon BTE is then solved analytically and numerically for SLBN, BLBN, 5LBN and Bulk-{\it h} BN to calculate $\kappa_L$ for a constant length over a wide range of temperatures and {\it vice versa} again in good agreement  with available experimental results.

Both the theoretical approaches, {\it i.e.} real space super cell and Callaway-Klemens, show the same magnitude for $\kappa_L$ but the temperature dependence by the two methods are different.
The lattice thermal conductivity for these materials are practically length independent for sample lengths greater than 100 $\mu$m which tends to their thermodynamic limit.
Calculated values for $\kappa_L$ for BLBN and 5LBN agree very well with experiments when calculated by the real space approach rather than by the Callaway-Klemens method. This may be because the experimental behavior of $\kappa_L$ for both BLBN and 5LBN tend to saturate at higher temperatures instead of having a maxima.
However, the Callaway-Klemens method agrees better with available experimental data for Bulk-{\it h} BN.
Further experiments could resolve this discrepancy.

Mode dependent numerical calculations using the Callaway-Klemens formalism suggest that ZA modes dominate only at very low temperatures and have the least contribution as the temperature is increased.
This is in stark conflict with our calculations based on real space super cell approach.
Since the velocities and Gr\"{u}neisen parameters are extremely similar for single and bi layer boron nitride, one would expect $\kappa_L$ for both the systems to be similar. However, in the case of graphene, we have a significant reduction in $\kappa_L$ which is seen both experimentally \cite{hongyang2014} and theoretically \cite{lindsay2011,RDSM17}. The larger $\kappa_L$ in SLBN in comparison to BLBN using the Callaway-Klemens method was due to the symmetry put by hand and not a consequence of the theory.
This implies that the closed form of the relaxation time used in Callaway-Klemens method is a poor approximation having little predictive value and one must solve the BTE beyond the RTA.
Our calculations suggests that for an enhanced figure of merit, $ZT$, in such materials, the sample length must be in the $\mu$m range or smaller and should be stacked on top of each other.


\section*{Acknowledgments}
We thank Dr. Jes\'{u}s Carrete of the Technical University, Vienna, for his insightful correspondence on the ShengBTE code for the calculation of the mode dependence contribution to the total thermal conductivity.
We also thank D.L. Nika of Moldova State University and A.A. Balandin of the University of California, Riverside, for their helpful correspondence based on their recent publications. All calculations were performed in the High Performance Cluster platform at the S.N. Bose National Centre for Basic Sciences. RD acknowledges support through a Senior Research Fellowship of the S.N. Bose National Centre for Basic Sciences.

\appendix*
\section {Derivation of Irreducible representations}

We define the reducible representation ($\Gamma_{\rm red}$) by placing three vectors on each atom in the unit cell which will obey the following rules when operated by a symmetry transformation.
\begin{itemize}
  \item If a vector is not moved (reversed) by an operation, it contributes 1 (-1) to $\chi$.
  \item If a vector is moved to a new location by an operation, it contributes 0 to $\chi$.
\end{itemize}
where $\chi$ is the character in the reducible representation.
Our reducible representations ($\Gamma_{red}$) are shown in the column before every new point group representation in table \ref{ct}. 
Using the reduction formula, $a_i = \frac{1}{g} \sum \chi_R \chi_{IR}$, where $a_i$ is the number of times an irreducible representation contributes to the reducible representation, $g$ is the total number of symmetry operations for a particular point group and $\chi_R$ ($\chi_{IR}$) is the corresponding character in the reducible (irreducible) representation, we derive the irreducible representations.

\begin{table*}[!]
\caption{\label{ct} The point group representation for SLBN, BLBN, 5LBN and Bulk-{\it h}BN at the $\Gamma$ point in the BZ. The irreducible representation is obtained from the reducible representation $\Gamma^X_{red}$ of the system X using the reduction formula.}
\begin{tabular}{lccccccccccccc}
\hline\hline
D$_{{\rm 6h}}$ & E & 2C$_{6}$ & 2C$_3$ & C$_2$ & 3C$_2^{'}$ & 3C$_2^{''}$ & i & 2S$_6$ & 2S$_3$ & $\sigma_{\rm h}$ & 3$\sigma_{\rm v}$ & 3$\sigma_{\rm d}$ & Basis \\
\hline
A$_{\rm 2u}$ & 1 & 1& 1 & 1& -1 & -1 & -1 & -1 &	-1 &	 -1& 1 & 1 & $z$ \\
B$_{\rm 1g}$	 & 1 & -1& 1 & -1 & 1 & -1 & 1 & 1 & -1 & -1 & -1 & 1 & $yz(3x^2-y^2)$ \\
B$_{\rm 2g}$	 & 1 & -1& 1 & -1 & -1 & 1 & 1 & 1 & -1 & -1 & 1 & -1 & $xz(x^2-3y^2)$ \\
E$_{\rm 2g}$ & 2 & -1 & -1 & 2 & 0 & 0 & 2 & -1 & -1 & 2 & 0 & 0 & \{$x^2-y^2,xy$\} \\
E$_{\rm 1u}$ & 2& 1& -1& -2 & 0 & 0 & -2 & 1 & -1 & 2 & 0 & 0 & \{$x,y$\} \\
\hline
$\Gamma^{bulk-{\it h}BN}_{\rm red}$ & 12 & 0 & 0 & 0 & 0 & -4 & 0 & -8 & 0 & 4 & 4 & 0 & \ \\
$\Gamma^{SLBN}_{\rm red}$ & 6 & 0 & 0 & 0 & -2 & 0 & 0 & -4 & 0 & 2 & 0 & 2 & \ \\
\hline
D$_{\rm 3d}$ & E & & 2C$_3$ & & 3C$_2^{'}$ & & i & & 2S$_6$ & & 3$\sigma_{\rm d}$ \\
\hline
A$_{\rm 2u}$	 & 1 & & 1 & & -1 & & -1 &  & -1 & & 1 & & $z$ \\
A$_{\rm 1g}$ & 1 & & 1 & & 1 & & 1 &  & 1 & & 1 & &  $z^2$ \\
E$_{\rm g}$	& 2 & & -1 & & 0 & & 2 &  & -1 & & 0 & &  \{$xz, yz$\} \\
E$_{\rm u}$	& 2 & & -1 & & 0 & & -2 &  & 1 & & 0 & &  \{$x, y$\} \\
\hline
$\Gamma^{BLBN}_{\rm red}$ & 12 & & 0 & & 4 & & 0 &  & 0 & & 0 & & \\
\hline
D$_{\rm 3h}$ & E & & 2C$_3$ & & 3C$_2^{'}$ & & $\sigma_{\rm h}$ & & 2S$_3$ & & 3$\sigma_{\rm v}$ \\
\hline
A$_1^{'}$	 & 1 & & 1 & & 1 & & 1 &  & 1 & & 1 & & $z^2$ \\
A$_2^{''}$ & 1 & & 1 & & -1 & & 1 &  & 1 & & -1 & &  $z$ \\
E$^{'}$	& 2 & & -1 & & 0 & & 2 &  & -1 & & 0 & &  \{$x, y$\} \\
E$^{''}$	& 2 & & -1 & & 0 & & -2 &  & 1 & & 0 & &  \{$xz, yz$\} \\
\hline
$\Gamma^{5LBN}_{\rm red}$ & 30 & & 0 & & 10 & & 2 &  & -4 & & -2 & & \\
\hline\hline
\end{tabular}
\end{table*}



%
\end{document}